\documentclass{article}

\usepackage[dvips]{graphicx}

\begin{document}
\begin{center}
{\Large Solitary pulses and periodic waves in the parametrically driven
complex Ginzburg-Landau equation} \bigskip

Hidetsugu Sakaguchi

Department of Applied Science for Electronics and Materials,
Interdisciplinary Graduate School of Engineering Sciences, Kyushu
University, Kasuga, Fukuoka 816-8580, Japan
\end{center}

\bigskip

\begin{center}
Boris Malomed

Department of Interdisciplinary Studies, Faculty of Engineering, Tel Aviv
University, Tel Aviv 69978, Israel

\bigskip

{\Large Abstract}
\end{center}
A one-dimensional model of a dispersive medium with intrinsic loss,
compensated by a parametric drive, is proposed. It is a combination of the
well-known parametrically driven nonlinear Schr\"{o}dinger (NLS) and
complex cubic Ginzburg-Landau equations, and has various physical
applications (in particular, to optical systems). For the case when the zero
background is stable, we elaborate an analytical approximation for
solitary-pulse (SP) states. The analytical results are found to be in good
agreement with numerical findings. Unlike the driven NLS equation, in the
present model SPs feature a nontrivial phase structure. Combining the
analytical and numerical methods, we identify a stability region for the SP
solutions in the model's parameter space. Generally, the increase of the
diffusion and nonlinear-loss parameters, which differ the present model from
its driven-NLS counterpart, lead to shrinkage of the stability domain.
At one border of the stability region,
the SP is destabilized by the Hopf bifurcation, which converts it into a
localized breather. Subsequent period doublings make internal vibrations of
the breather chaotic. In the case when the zero background is unstable,
hence SPs are irrelevant, we construct stationary periodic solutions, for
which a very accurate analytical approximation is developed too. Stability
of the periodic waves is tested by direct simulations.

\newpage

\section{Introduction}

One of paradigm models adopted for the description of pulse dynamics in
nonlinear dispersive media with intrinsic loss is the parametrically driven
damped nonlinear Schr\"{o}dinger (NLS) equation. In the one-dimensional
case, it is 
\begin{equation}
iu_{t}+\frac{1}{2}u_{xx}+\frac{1}{2}|u|^{2}u=(k-i)u+\gamma u^{\ast },
\label{NLS}
\end{equation}
where $u(x,t)$ is the complex wave field, the asterisk stands for the
complex conjugation, real $k$ is the frequency mismatch of the parametric
drive, real $\gamma $ is the drive's amplitude, which may be assumed
positive, and the loss parameter, as well as the ones accounting for the
spatial diffraction/dispersion and nonlinear frequency shift, are normalized
to be $1$. Equation (\ref{NLS}) finds applications to various physical
problems, such as dynamics of spin waves in magnetics \cite{magnetic},
plasma waves \cite{plasma}, and Faraday surface waves in a fluid layer
driven by a time-periodic external force \cite{Faraday}. Arguably, the most
realistic application of eq.~(\ref{NLS}) is to models of nonlinear optics,
such as a cavity filled with a Kerr-nonlinear lossy material, the driving
term being provided for by an external field supplied at the frequency
coinciding with that of the signal field, see, e.g., ref.~4 and
references therein.

It is well known that eq.~(\ref{NLS}) gives rise to two exact solutions for
phase-locked solitons\cite{Miles}, 
\begin{equation}
u(x)=A(x)e^{i\theta },\,A(x)=2\sqrt{k+\gamma \cos 2\theta }\,{\rm sech}
\left( \sqrt{2(k+\gamma \cos 2\theta )}x\right) ,  \label{soliton}
\end{equation}
where the phase constant takes values 
\begin{equation}
\theta _{1}=-\frac{1}{2}\sin ^{-1}\left( 1/\gamma \right) ,\,\theta _{2}=
\frac{1}{2}\left[ \pi +\sin ^{-1}\left( 1/\gamma \right) \right] \,.
\label{theta}
\end{equation}
The solitons exist provided that $\gamma >1$.

The soliton with $\theta =\theta _{2}$, corresponding to the smaller value
of the amplitude, $A_{0}\equiv 2\sqrt{k+\gamma \cos 2\theta }$, is always
unstable, while the other one may be stable. The soliton stability in this
model was studied in detail \cite{Barash1}. In particular, an obvious
necessary condition is stability of the zero solution, which implies 
\begin{equation}
k>\sqrt{\gamma ^{2}-1}.  \label{zerostability}
\end{equation}
Thus, the stable soliton may exist in the interval $1<\gamma ^{2}<1+k^{2}$
of the values of gain. Actually, the soliton is stable only in a part of
this region, while in another part it is destabilized via a Hopf
bifurcation, which was also studied in detail, see an earlier work \cite
{Okamura} and a systematic analysis in refs.~8.

Another important dynamical model for pattern formation in nonlinear lossy
media is based on the complex cubic Ginzburg-Landau (CCGL) equation, which
may be written as 
\begin{equation}
iu_{t}+(\beta _{1}-i\beta _{2})u_{xx}+(c_{1}+ic_{2})|u|^{2}u=(k+i)u,
\label{GL}
\end{equation}
where the coefficients $\beta _{1,2}$ and $c_{1,2}$ are real, and $\beta _{2}
$ and $c_{2}$ cannot be negative. The linear-gain coefficient on the
right-hand side of eq.~(\ref{GL}) is normalized to be $1$ [cf. the
normalized linear-loss coefficient in eq.~(\ref{NLS})]. The CCGL equation
has many applications, including description of the onset of turbulence in
the Poiseuille flow \cite{Poiseuille} and Langmuir waves in plasmas \cite
{Lennart}. Additionally, if $t$ and $x$ are replaced, respectively, by the
propagation distance $z$ and reduced time $\tau $, eq.~(\ref{GL}) describes
the propagation of light signals in a pumped nonlinear optical waveguide,
with the dispersion coefficient $\beta _{1}$, Kerr coefficient $c_{1}$,
filtering coefficient $\beta _{2}$, and coefficient of two-photon absorption 
$c_{2}$ \cite{HK}.

Various solutions to eq.~(\ref{GL}) and their physical applications have
been studied in detail, see a recent review \cite{Aranson}. A particular
solution, which is unstable but nevertheless quite meaningful (see, e.g., a
detailed consideration in ref.~13), is a solitary pulse (SP) that
can be found in an exact analytical form \cite{Poiseuille,Lennart} 
\begin{equation}
u=A\,_{0}\left[ {\rm sech}\left( \kappa x\right) \right] ^{1+i\mu }\exp
\left( -i\omega t\right) ,  \label{pulse}
\end{equation}
where all the real constants $A_{0},\kappa ,\mu $ and $\omega \,$\ are
uniquely expressed in terms of parameters of eq.~(\ref{GL}). It is
noteworthy that the NLS equation proper, and its damped parametrically
driven version (\ref{NLS}), admit soliton solutions only under the condition
which, in terms of eq.~(\ref{GL}), is $\beta _{1}c_{1}>0$ (in terms of
nonlinear optics, it implies that the nonlinear waveguide is either
self-focusing with anomalous group-velocity dispersion, or self-defocusing
with normal dispersion). However, except for the case $\beta _{2}=c_{2}=0$,
the general CCGL equation (\ref{GL}) has the exact SP solution in the form 
(\ref{pulse}) irrespective of the sign of the product $\beta _{1}c_{1}$, the
only additional condition being that $\beta _{1}$ and $c_{1}$ must not
simultaneously vanish. The most important difference of the SP solution (\ref
{pulse}) from the phase-locked soliton (\ref{soliton}) is the presence of
the {\it chirp} $\mu$, which accounts for a nontrivial intrinsic phase
structure in the SP.

A natural generalization of both models (\ref{NLS}) and (\ref{GL}) is a
parametrically driven CCGL equation, which we take in the form

\begin{equation}
iu_{t}+(\beta _{1}-i\beta _{2})u_{xx}+(c_{1}+ic_{2})|u|^{2}u=(k-i)u+\gamma
u^{\ast }.  \label{model}
\end{equation}
It may apply to the above-mentioned physical systems under properly modified
conditions. For instance, in the case of the nonlinear optical waveguide,
eq.~(\ref{model}) describes a situation when the loss is compensated not by
intrinsic gain, but rather by a co-propagating pump wave. It should be
mentioned that a parametrically driven model of that type was already
introduced in a two-dimensional situation, which corresponds to the light
propagation in a planar nonlinear waveguide \cite{we}, and stable anisotropic
two-dimensional SPs were found. In one dimension, a CCGL system with a
nonlocal parametric drive was considered recently \cite{Knobloch}, but the
straightforward one-dimensional model (\ref{model}) has not been considered
yet. Complex dynamics of interfaces was also studied using this type equation
\cite{Mizuguchi}.

The existence of the stable soliton (\ref{soliton}) in the driven-NLS limit
of the model (\ref{model}), and of the unstable SP (\ref{pulse}) in its CCGL
counterpart, suggest a possibility to find a SP solution to eq.~(\ref{model}) and investigate its stability, 
which is the subject of the present work.
One may expect that an essential difference of the SP solution from its NLS
phase-locked counterpart (\ref{soliton}) is a nontrivial intrinsic phase
structure (chirp).

In section 2, we report results obtained for SPs in the model (\ref{model}).
First, we develop an analytical approximation, that predicts all the
properties of SPs, except for the chirp, with good accuracy. Then, we
demonstrate numerical results, which are summarized in the form of stability
diagrams in the model's phase space. The stability domain shrinks with the
increase of the diffusion and nonlinear-loss parameters $\beta _{1}$ and 
$c_{1}$ in eq.~(\ref{model}), which make it different from the NLS
counterpart (\ref{NLS}). At one of the domain's borders, the soliton is
destabilized via a Hopf bifurcation, which gives rise to a stable breather.
The breathers are also briefly considered in section 2.

In the case when the condition (\ref{zerostability}) for the stability of
the zero solution is not met, SPs keep to exist, but they cannot be stable.
In this case, relevant solutions are periodic waves. We briefly consider
them in section 3, also first in an approximate analytical form, and then by
means of numerical methods. In this case, analytical and numerical results
are virtually indistinguishable.

\section{Solitary-pulses}

\subsection{Analytical approximation}

To develop an analytical approximation to SP solutions, we fix two
parameters in eq.~(\ref{model}), $\beta _{1}=c_{1}=1/2$, which can always
be done by means of rescaling (provided that $\beta_{1}$ and $c_{1}$ are
positive). We then assume an approximate form for the solution which mimics
the exact solution (\ref{soliton}) available in the NLS limit (\ref{NLS}), 
\begin{equation}
u(x)=A(x)e^{i\theta},  \label{ansatz}
\end{equation}
with {\em constant} $\theta $. Then, substituting the ansatz (\ref{ansatz})
into eq.~(\ref{model}), we formally obtain two different equations for the
single real function $A(x)$: 
\begin{equation}
\frac{d^{2}A}{dx^{2}}=2\left[ k+\gamma \cos (2\theta )\right] A-A^{3},
\label{A1}
\end{equation}
\begin{equation}
\beta _{2}\frac{d^{2}A}{dx^{2}}=\left[ 1+\gamma \sin (2\theta )\right]
A+c_{2}A^{3}.  \label{A2}
\end{equation}
We consider the case that $\beta_2,\,1+\gamma\sin(2\theta)$ and $c_2$ are small. Eliminating the derivative term from eqs.(\ref{A1}) and (\ref{A2}), we
arrive at a simpler relation, 
\begin{equation}
(c_{2}+\beta _{2})A^{3}=\left( 2\beta _{2}\left[ k+\gamma \cos \left(
2\theta \right) \right] -\left[ 1+\gamma \sin (2\theta )\right] \right) A.
\label{simple}
\end{equation}
An exact solution to eq.~(\ref{A1}) is given by the expression for $A(x)$
from eq.~(\ref{soliton}), which we will use here, but without adopting the
relation (\ref{theta}). Instead, we substitute the expression for $A(x)$
into eq.~(\ref{simple}), then multiply all the equation again by $A(x)$, and
integrate the result over $x$ from $-\infty $ to $+\infty $. This procedure
leads to the following relation, that we use as a definition of $\theta $ in
the waveform (\ref{soliton}), which is now considered as an approximate
solution to eq.~(\ref{model}): 
\begin{equation}
2(k+\gamma \cos 2\theta )(4c_{2}+\beta _{2})+3\left[ 1+\gamma \sin \left(
2\theta \right) \right] =0\,.  \label{approximation}
\end{equation}
In particular, eq.~(\ref{approximation}), with regard to eq.~(\ref
{zerostability}), predicts that a stable SP solution may exist in the
interval 
\begin{equation}
\sqrt{\gamma ^{2}-1}<k<k_{{\rm cr}}\equiv \frac{\gamma \sqrt{
9+4(4c_{2}+\beta _{2})^{2}}-3}{2\left( 4c_{2}+\beta _{2}\right) }\,.
\label{k_max}
\end{equation}
With the increase of the combination $\left( 4c_{2}+\beta _{2}\right)$, 
$k_{{\rm cr}}$ attains a minimum value, 
\begin{equation}
\left( k_{{\rm cr}}\right) _{\min }=\sqrt{\gamma ^{2}-1},  \label{minimum}
\end{equation}
at 
\begin{equation}
4c_{2}+\beta _{2}=\left( 3/2\right) \sqrt{\gamma ^{2}-1}  \label{min}
\end{equation}
[at this point, the interval (\ref{k_max}) collapses into a point]. With the
further increase of $\left( 4c_{2}+\beta _{2}\right) $, $k_{{\rm cr}}$ grows
up to a maximum value, $\left( k_{{\rm cr}}\right) _{\max }\equiv k_{{\rm cr}
}\left( 4c_{2}+\beta _{2}=\infty \right) =\gamma .$
\begin{figure}[htb]
\begin{center}
\includegraphics[width=12cm]{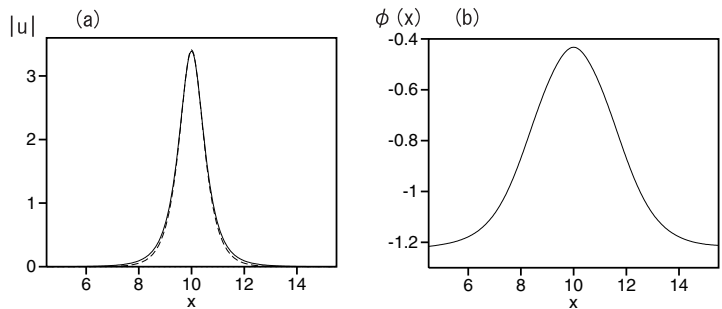}
\caption{A typical example of the stable solitary pulse. The panels (a) and
(b) show the amplitude and phase profiles of the stationary solution with 
$\gamma =1.5,k=2$ and $\beta _{2}=c_{2}=0.02$. In (a), the solid and dashed
curves represent the numerical result and analytical approximation [based on
Eqs. (\ref{soliton}) and (\ref{approximation})], respectively.}
\label{fig:1} 
\end{center}
\end{figure} 
\begin{figure}[htb]
\begin{center}
\includegraphics[width=12cm]{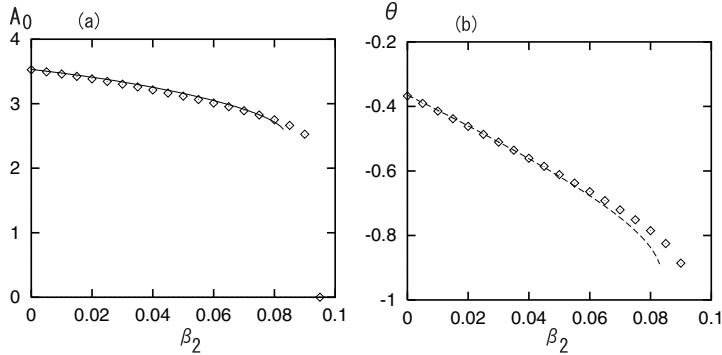}
\caption{The amplitude of the solitary pulse (a) and its average phase (b),
as defined in Eq. (\ref{averphase}), vs. the parameter $\beta _{2}=c_{2}$.
Other coefficients in Eq. (\ref{model}) are fixed: $\gamma =1.5,k=2$. The
symbols show numerical results, while continuous curves are predicted by the
analytical approximation.}
\label{fig:2} 
\end{center}
\end{figure} 

\begin{figure}[htb]
\begin{center}
\includegraphics[width=7cm]{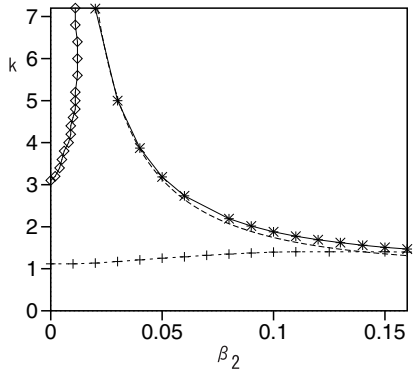}
\caption{The region in the parametric plane with the coordinates $\beta
_{2}=c_{2}$ and $k$ ($\gamma =1.5$ is fixed) in which stable solitary-pulse
solutions were found. The curves connecting symbols which mark
stability/existence borders are guides for the eye. The dashed curve, close
to the upper right border, is the analytical prediction (\ref{k_max}).}
\label{fig:3} 
\end{center}
\end{figure} 

\begin{figure}[htb]
\begin{center}
\includegraphics[width=7cm]{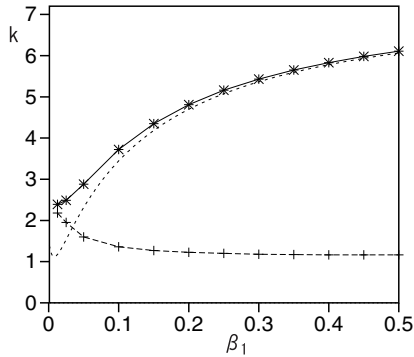}
\caption{The solitary-pulse stability region in the
parameter plane ($\beta _{1},k$) at fixed values $c_{1}=0.5,\beta
_{2}=c_{2}=0.025$, and $\gamma =1.5$. The pulses are stable between the two
borders. The dashed curve, close 
to the upper right border, is the analytical prediction.}
\label{fig:4} 
\end{center}
\end{figure}
 
\begin{figure}[htb]
\begin{center}
\includegraphics[width=14cm]{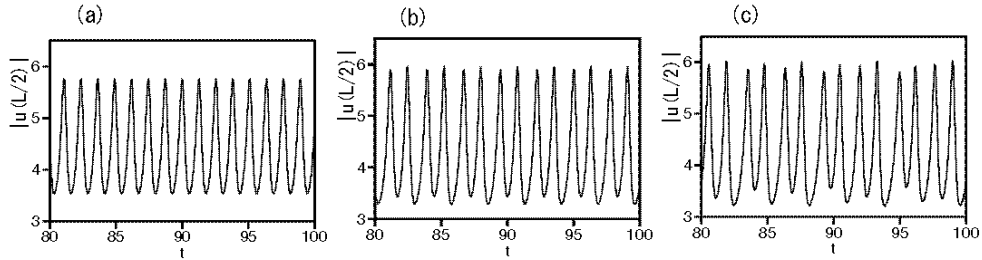}
\caption{Time evolutions of the breather's
amplitude at (a) $\beta_2=c_2=0.005$, (b) $\beta_2=c_2=0.002$ and (c) $\beta_2=c_2=0.001$ in the case $k=4.5,c_{1}=\beta _{1}=0.5,\gamma =1.5$.}
\label{fig:5} 
\end{center}
\end{figure} 
\begin{figure}[htb]
\begin{center}

\includegraphics[width=7cm]{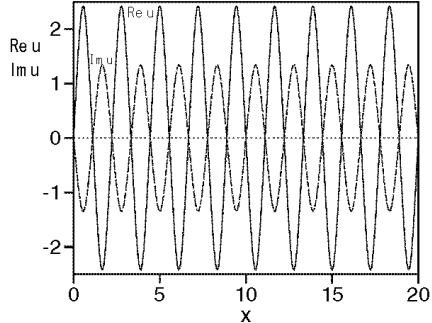}
\caption{Comparison of the real and imaginary parts of the approximate
analytical solution (\ref{cn}), i.e., $A_{0}{\rm cn}\left( \kappa x,q\right)
\cos \theta $ (solid curve) and $A_{0}{\rm cn}\left( \kappa x,q\right) \sin \theta $ (dashed curve), and a direct numerical solution, in the case
with $c_{2}=\beta _{2}=0.02$, $k=-2,\gamma =1.5$, and the period $\Lambda
=20/9$. Actually, the analytical curves completely overlap with the
numerical ones. This solution is dynamically stable.}
\label{fig:6} 
\end{center}
\end{figure}
 
\begin{figure}[htb]
\begin{center}
\includegraphics[width=12cm]{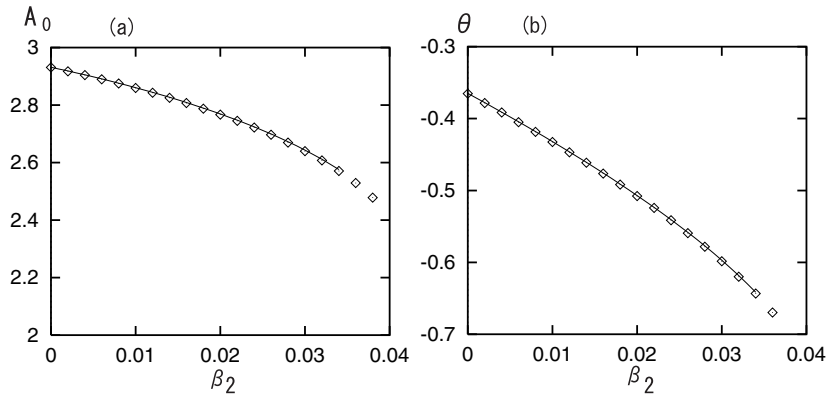}
\caption{Comparison of the amplitude (a) and average phase (b) of the stable
stationary periodic solutions, as found from the analytical approximation
(continuous curves) and numerical solution (symbols). The dependences are
obtained for the case $\beta _{2}=c_{2}$, the other parameters being fixed,
with $k=-2$, $\gamma =1.5$, and the period $\Lambda =20/9$.}
\label{fig:7} 
\end{center}
\end{figure}
 
\begin{figure}[htb]
\begin{center}
\includegraphics[width=7cm]{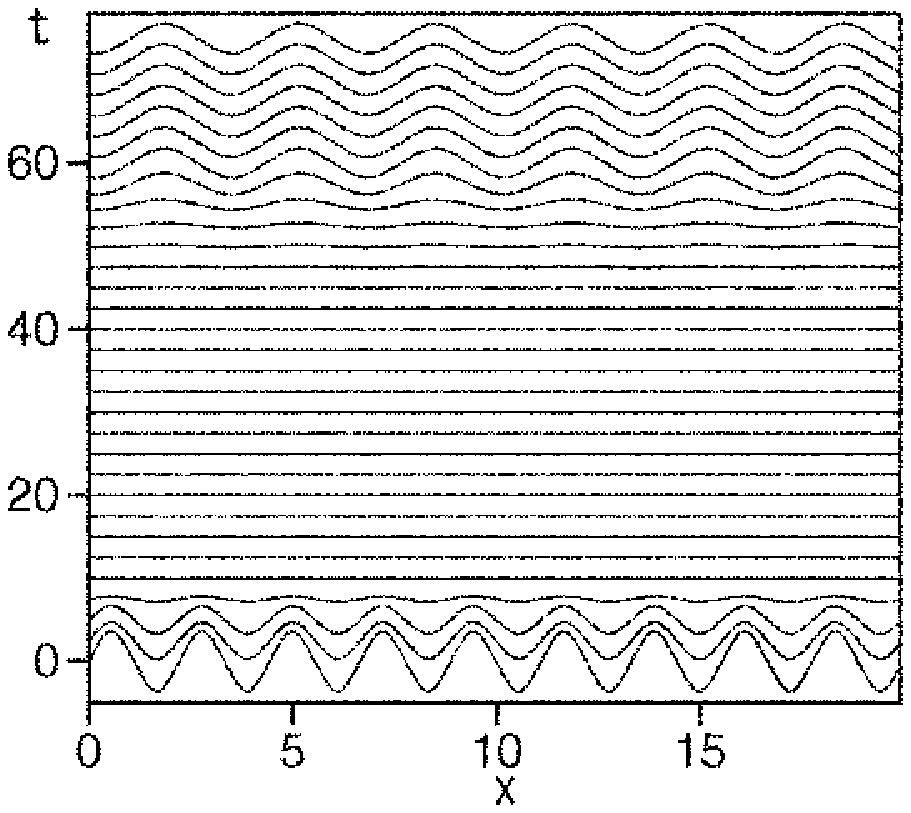}
\caption{Time evolution of the real part of $u$ in the case with $c_2=\beta_2=0.045$, $k=-2,\gamma=1.5$. The period is $\Lambda=20/9$ in the initial state, and the period is $\Lambda=20/6$ in the final state.}
\label{fig:8} 
\end{center}
\end{figure} 
\subsection{Numerical results}

The stationary version of eq.~(\ref{model}) was solved numerically, and the
findings were compared to those obtained in the approximate analytical form
based on eqs.~(\ref{soliton}) and (\ref{approximation}). A typical example
is displayed in Fig.~1, where the panel (a) shows the numerically found
profile of $|u(x)|$ together with the analytical approximation to it, and
the panel (b) is the corresponding phase profile, which was found only in
the numerical form, as the analytical approximation assumes a constant phase.

Results for the stationary SP solutions can be summarized in the form of the
plot displayed in Fig.~2. Its panel (a) shows the soliton's amplitude, as
found numerically and in the analytical approximation, vs. the coefficient 
$\beta _{2}=c_{2}$, which measures the deviation of the model (\ref{model})
from the NLS counterpart (\ref{NLS}), and the panel (b) shows the average
value of the SP's phase $\phi (x)$, which is defined as 
\begin{equation}
\theta \equiv \frac{\int_{-\infty }^{+\infty }\phi (x)|u(x)|^{2}dx}
{\int_{-\infty }^{+\infty }|u(x)|^{2}dx},  \label{averphase}
\end{equation}
in comparison with the analytical prediction for the average phase, which is
realized as the phase $\theta $ defined by eq.~(\ref{approximation}). The
numerical and analytical results displayed in Fig.~2 terminate at points
where, respectively, the numerical solution and eq.~(\ref{approximation})
cease to produce any nontrivial result (the numerical solution shows that,
beyond the termination point, only the trivial state, $u=0$, can be found).

The results are further summarized in the diagram shown in Fig.~3, which
displays a region in the parameter plane ($\beta _{2}=c_{2},k$), in which 
{\em stable} stationary SP solutions were found [the stability was verified
by direct simulations of the full time-dependent equation (\ref{model})].
The upper right border marked by stars is the one beyond which only the zero
solution can be found numerically; the dashed curve is an analytical
prediction for the same border as given by eq.~(\ref{k_max}), i.e., the
boundary beyond which eq.~(\ref{approximation}) yields no solutions. Below
the lower border, which is marked by crosses, stable SPs cannot be found. 
The width of an initial pulse increases and the pulse spreads out to a spatially extended state. (The spatially extended state is not a uniform state.) The lower border is slightly upper than the instability line $k=\sqrt{\gamma^2-1}$ of the zero solution.  The stability region in Fig.~3 ends approximately at  $\left( \beta _{2}\right) _{{\rm cutoff}}\approx
0.16$, where the upper border approaches the lower border.  
Lastly, the upper left border, marked by rhombuses, is a line of the Hopf
bifurcation destabilizing the SP: behind this border, a stable solution is a
regularly or chaotically vibrating breather, see below.

The consideration of the above results, and the comparison with the
analytical prediction shows that, first, the analytical approximation turns
out to be fairly accurate, and, second, the stability region quickly shrinks with the
increase of the diffusion and nonlinear-loss coefficients $\beta _{2}$ and 
$c_{2}$ in eq.~(\ref{model}). The stability region in Fig.~3 greatly
extends in the direction of larger values of $k$ at small finite values of 
$\beta _{2}=c_{2}$ [which may be easily explained by the analytical result
showing that the maximum value of $k$, as given by eq.~(\ref{k_max}),
diverges as $\beta _{2}=c_{2}\rightarrow 0$].

Another relevant representation of the SP stability region is in terms of
the parametric plane ($\beta _{1},k$) for fixed values $\beta_2=c_2=0.025,\,c_1=0.5$ and $\gamma=1.5$ of the other parameters, which is shown in Fig. 4. 
The two borders in Fig.~4 denote the same types of bifurcation curves as in Fig.~3.
The upper  border marked by stars is the one beyond which only the zero
solution can be found numerically; the dashed curve is an analytical
prediction $k_{{\rm cr}}= \{\gamma \sqrt{
9+(8c_{2}+\beta _{2}/\beta_1)^{2}}-3\}/( 8c_{2}+\beta _{2}/\beta_1)$, which is a modified one of eq.~(\ref{k_max}).
The analytical prediction is good for large $\beta_1$, however, 
the approximation breaks down for small $\beta_1$, since the derivation of  Eq. (\ref{k_max}) is based on the conditions  $\beta_2\ll\beta_1$ and $c_2\ll c_1$. 
Below the lower border, which is marked by crosses, stable SPs cannot be found.   Note that the (unstable) exact SP
solution (\ref{pulse}) to the usual CCGL equation exists, unless both $\beta
_{2}$ and $c_{2}$ vanish, not only at positive but also at negative values
of the product $\beta _{1}c_{1}$ \cite{Lennart}, which is a drastic
difference from the classical NLS solitons that exist solely in the case 
$\beta _{1}c_{1}>0$. Nevertheless, Fig.~4 shows that the parametrically
driven CCGL supports, as well as its NLS counterpart, (stable) SPs only in
the case $\beta _{1}c_{1}>0$. 

\subsection{Breathers}

As it was mentioned above, the SP solution suffers destabilization via the
Hopf bifurcation at the border marked by rhombuses in Fig.~3, similar to
what is known about solitons in the usual driven-NLS model \ref{NLS} \cite
{Okamura,Barash1,Barash2}. The Hopf bifurcation gives rise to stable
oscillatory pulses in the form of breathers. Similar breathing pulses are also found in the quintic CCGL equation \cite{Brand}. While we did not aim to
investigate the breathers in detail, we have found that they always keep a
single-humped shape. With further variation of the parameters, the breather
performing regular oscillations undergoes period-doubling bifurcations,
which do not destabilize the pulse as a whose, but render its intrinsic
oscillations chaotic. Figure 5 shows the time evolutions of 
the breather's amplitude, which is equal to $|u(L/2)|$, for $\gamma=1.5$ and $k=4.5$.  The parameters $\beta_1$ and $c_1$ are again fixed to be 0.5.
The amplitude of the pulse exhibits a limit cycle oscillation 
at $\beta_2=c_2=0.005$, a limit cycle oscillation with doubled period is seen at $\beta_2=c_2=0.002$, 
and  the time evolution becomes chaotic at $\beta_2=c_2=0.001$.  

\section{Periodic waves}

In the case when the condition (\ref{zerostability}) does not hold and the
zero solution is unstable, SP solutions are irrelevant. However, it is
natural to expect that the model supports spatially periodic stationary
waves in this case, which may be quite relevant in the corresponding
physical settings, provided that these waves are stable.

Periodic solutions can be first analyzed by means of an analytical
approximation, similar to what was done above for the SP solutions, i.e.,
adopting the ansatz (\ref{ansatz}) with constant $\theta $. Then, an exact
solution to the corresponding equation (\ref{A1}) is 
\begin{equation}
u(x)=A_{0}\,{\rm cn}(\kappa x,q)e^{i\theta },  \label{cn}
\end{equation}
where ${\rm cn}$ is the elliptic cosine with the modulus $q=A_{0}/\sqrt{
a^{2}+A_{0}^{2}}$, and other constants are defined as follows: $\kappa =
\sqrt{(a^{2}+A_{0}^{2})/2}$, $A_{0}^{2}=2(k+\gamma \cos 2\theta )+\sqrt{
4(k+\gamma \cos 2\theta )+2E}$, and $a^{2}=-2(k+\gamma \cos 2\theta )+\sqrt{
4(k+\gamma \cos 2\theta )+2E}$. In these expressions, $E$ is a free
parameter of the family of the periodic solutions, which can be fixed if a
certain period $\Lambda $ is selected by periodic conditions. In terms of
the above parameters, 
\begin{equation}
\Lambda =4\sqrt{2}K(q)/\sqrt{a^{2}+A_{0}^{2}},  \label{period}
\end{equation}
\newline
where $K(q)$ is the complete elliptic integral of the first kind.

To determine the value of $\theta $, we follow the same approach as in the
case of SP, viz., we substitute the expression (\ref{cn}) into Eq. (\ref
{simple}), then multiply it again by the expression (\ref{cn}), and
integrate over the period (\ref{period}). The resultant equation was solved
numerically, in order to find $\theta $ for a given period $\Lambda $.

In a direct numerical simulation, we have obtained a stable pattern with the period $\Lambda=20/9$ at $c_{2}=\beta_{2}=0.02$, $k=-2$ and $\gamma =1.5$.
In Fig.~6, we compare the direct numerical solution and the predicted approximate analytical solution for the periodic wave. 
A more general comparison
of the analytical and numerical results is presented in Fig.~7, cf. a
similar picture for the SP solutions in Fig.~2 [the average phase shown in
Fig.~7(b) is defined by the expression (\ref{averphase}), in which the
integration is limited to one period of the solution; in fact, the phase of
the periodic solution is almost constant, unlike that of SPs]. The
dependences in Fig.~7 terminate at points beyond which the numerical solutions could not be continued. As concerns the
numerical solution, in the case shown in Fig.~7,  a solution with $\Lambda=20/9$ cannot be found for $\beta _{2}>0.04$. 
Figure 8 displays a time evolution at $\beta_2=c_2=0.045$. 
The initial condition is  the periodic solution with the period $\Lambda=20/9$ 
 for $\beta_2=c_2=0.02$ shown in Fig.~6. The amplitude of the periodic solution with the period $\Lambda=20/9$ decays to zero, and another periodic solution 
with period $\Lambda=20/6$ appears as a stable solution.

\section{Conclusion}

In this work, we have introduced a model of the one-dimensional nonlinear
dispersive medium with intrinsic loss, which are compensated by a parametric
driving field. The model is a combination of the parametrically driven
nonlinear Schr\"{o}dinger (NLS) and complex cubic Ginzburg-Landau equations,
and has various physical applications (in particular, to optical systems).
In the case when the zero background is stable, we developed an analytical
approximation for a solitary-pulse (SP) solution, which was found to be in
good agreement with direct numerical results. Unlike the well-studied case
of the driven damped NLS equation, in the present case SPs have a nontrivial
phase structure. Combining the analytical and numerical methods, we have
identified the stability domain for the SP solutions. Generally, the
increase of the diffusion and nonlinear-loss parameters leads to shrinkage
of the stability region, but a nontrivial feature is that it may stretch in
the direction of large values of the frequency mismatch. At one border of
the stability domain, the SP is destabilized by the Hopf bifurcation, which
gives rise to a localized breather. Further period doublings make internal
vibrations of the breather chaotic. In the case when the zero background is
unstable, hence SPs are irrelevant, we constructed stationary periodic
solutions, for which a very accurate analytical approximation was developed,
and their stability was studied in direct simulations.


\begin{thebibliography}{99}
\bibitem{magnetic}  H. Yamazaki and M. Mino, Progr. Theor. Phys. Suppl. {\bf
98} (1989) 400.

\bibitem{plasma}  V.E. Zakharov, S.I. Musher, and A.M. Rubenchik, Phys. Rep. 
{\bf 129} (1985) 185; N. Yajima and M. Tanaka, Progr. Theor. Phys. Suppl. 
{\bf 94} (1988) 138.

\bibitem{Faraday}  J. Wu, R. Keolian, and I. Rudnick, Phys. Rev. Lett. {\bf
52} (1984) 1421; J. Miles and D. Henderson, Ann. Rev. Fluid Mech. {\bf 22},
 (1990) 143.

\bibitem{cavity}  W.J. Firth, G.K. Harkness, A. Lord, J.M. McSloy, and D.
Gomila, J. Opt. Soc. Am. B {\bf 19} (2002) 747.

\bibitem{Miles} J.W. Miles, J. Fluid Mech. {\bf 148} (1984) 451.

\bibitem{Barash1}  I.V. Barashenkov, M. M. Bogdan, and V.I. Korobov,
Europhys. Lett. {\bf 15} (1991) 113.

\bibitem{Okamura}  A. Okamura and H. Konno, J. Phys. Soc. Jpn. {\bf 58},
 (1989) 1930.

\bibitem{Barash2}  M. Bondila, I.V. Barashenkov, and M.M. Bogdan, Physica D 
{\bf 87} (1995) 314; N.V. Alexeeva, I.V. Barashenkov, and D.E. Pelinovsky,
Nonlinearity {\bf 12} (1999) 103.

\bibitem{Poiseuille}  L.M.~Hocking and K.~Stewartson, Proc. Roy. Soc. London 
{\bf 326,}  (1972) 289.

\bibitem{Lennart}  N.R.~Pereira and L. Stenflo, Phys. Fluids {\bf 20,} 
(1977) 1733.

\bibitem{HK}  A. Hasegawa and Y. Kodama. {\it Solitons in Optical
Communications} (Oxford University Press:\ Oxford, 1995).

\bibitem{Aranson}  I.S. Aranson and L. Kramer, Rev. Mod. Phys. {\bf 74} 
(2002) 99.

\bibitem{Jena}  B.A. Malomed, M. G\"{o}lles, I.M. Uzunov, and F. Lederer,
Physica Scripta {\bf 55} (1997) 73.

\bibitem{we}  H. Sakaguchi and B.A. Malomed, Physica D {\bf 167} (2002) 123.

\bibitem{Knobloch}  M. Higuera, J. Porter, and E. Knobloch, Physica D {\bf
162} (2002) 155.

\bibitem{Mizuguchi} T. Mizuguchi and S. Sasa, Prog. Theor. Phys. {\bf 89} (1993) 599. 

\bibitem{Brand} R.J. Deissler and H.R. Brand, Phys. Rev. Lett. {\bf 72} (1994) 478.
\end{thebibliography}
\end{document}